
\overfullrule=0pt

\input harvmac
\input epsf


\def\a{{\alpha}}
\def\l{{\lambda}}
\def\b{{\beta}}
\def\g{{\gamma}}
\def\d{{\delta}}

\def\half{{1\over 2}}
\def\p{{\partial}}
\def\pb{{\bar\partial}}
\def\t{{\theta}}
\def\hat{\widehat}
\def\bar{\overline}

\def\ah{{\widehat\alpha}}
\def\lh{{\widehat\lambda}}
\def\bh{{\widehat\beta}}
\def\gh{{\widehat\gamma}}
\def\dh{{\widehat\delta}}

\def\o{{\omega}}
\def\oh{{\widehat\omega}}
\def\th{{\widehat\theta}}
\def\L{{\Lambda}}
\def\CH{{\cal H}}
\def\CO{{\cal O}}
\def\ll{{\langle}}
\def\rr{{\rangle}}


\lref\nv{N. Berkovits and C. Vafa,{ \it $N=4$ Topological Strings}
, Nucl. Phys. B433 (1995) 123, hep-th/9407190.}

\lref\csm{W. Siegel, {\it Classical Superstring Mechanics},
Nucl. Phys. B263 (1986) 93.}

\lref\bigpic{N. Berkovits, M.T. Hatsuda and W. Siegel, {\it The Big Picture},
Nucl. Phys. B371 (1992) 434, hep-th/9108021.}

\lref\berk{N. Berkovits, {\it Super-Poincar\'e
Covariant Quantization of the Superstring,} JHEP 04 (2000) 018,
hep-th/0001035.}

\lref\calabi{N. Berkovits, {\it Covariant Quantization of the
Green-Schwarz Superstring in a Calabi-Yau Background,}
Nucl.Phys. B431 (1994) 258, hep-th/9404162.}

\lref\adswitten{N. Berkovits, C. Vafa and E. Witten,
{\it Conformal Field Theory
of $AdS$ Background with Ramond-Ramond Flux,} JHEP 9903 (1999)
018, hep-th/9902098.}

\lref\bersha{M. Bershadsky, S. Zhukov and A. Vaintrop, {\it $PSL(N | N)$
Sigma Model as a Conformal Field Theory, } Nucl.Phys. B559 (1999) 205,
hep-th/9902180.}

\lref\bz{N. Berkovits, M. Bershadsky, T. Hauer, S. Zhukov and B. Zwiebach,
{\it Superstring Theory on $AdS_2\times S^2$ as a Coset Supermanifold,}
Nucl. Phys. B567 (2000) 61, hep-th/9907200.}

\lref\six{N. Berkovits, {\it Quantization of the
Type II Superstring in a Curved Six-Dimensional Background,}
Nucl. Phys. B565 (2000) 333, hep-th/9908041.}

\lref\metsaev{R. Metsaev and A. Tseytlin, {\it Type
IIB Superstring Action in $AdS_5\times S^5$ Background,}
Nucl. Phys. B533 (1998) 109, hep-th/9805028.}

\lref\cohom{N. Berkovits, {\it Cohomology in the Pure Spinor
Formalism for the Superstring,} JHEP 0009 (2000) 046, hep-th/0006003.}

\lref\dw{L. Dolan and E. Witten, {\it Vertex Operators for
$AdS_3$ Background with Ramond-Ramond Flux,} JHEP 9911 (1999) 003,
hep-th/9910205 .}

\lref\val{N. Berkovits and B.C. Vallilo,
{\it Consistency of Super-Poincar\'e Covariant Superstring
Tree Amplitudes,} JHEP 07 (2000) 015, hep-th/0004171.}

\lref\chan{N. Berkovits and O. Chand\'{\i}a, {\it Superstring Vertex
Operators in an $AdS_5\times S^5$ Background,} Nucl. Phys. B596 (2001)
185, hep-th/0009168.}

\lref\massive{N. Berkovits and O. Chand\'{\i}a, {\it Massive Superstring
Vertex Operator in D=10 Superspace,} hep-th/0204121.}

\lref\sugra{N. Berkovits and P. Howe, {\it Ten-Dimensional Supergravity
Constraints from the Pure Spinor Formalism for the Superstring,}
Nucl.Phys. B635 (2002) 75, hep-th/0112160.}

\lref\wave{N. Berkovits, {\it Conformal Field Theory for the
Superstring in a Ramond-Ramond Plane Wave Background,}
JHEP 0204 (2002) 037, hep-th/0203248.}

\lref\boer{J. de Boer and K. Skenderis, {\it Covariant Computation of the
Low Energy Effective Action of the Heterotic Superstring,}
Nucl.Phys. B481 (1996) 129, hep-th/9608078.}

\lref\adscft{J. Maldacena, {\it The Large N Limit of Superconformal Field
Theories and Supergravity,} Adv.Theor.Math.Phys. 2 (1998) 231,
hep-th/9711200 ; S. S. Gubser, I. R. Klebanov and A. M. Polyakov,
{\it Gauge Theory Correlators from Non-Critical String Theory,}
Phys.Lett. B428 (1998) 105, hep-th/9802109 ; E. Witten,
{\it Anti De Sitter Space And Holography,} Adv.Theor.Math.Phys. 2 (1998) 253,
hep-th/9802150.}

\lref\adsrev{O. Aharony, S. S. Gubser, J. Maldacena, H. Ooguri and Y. Oz,
{\it Large N Field Theories, String Theory and Gravity,}
Phys.Rept. 323 (2000) 183, hep-th/9905111.}

\lref\zamo{A. B. Zamolodtchikov, {\it 'Irreversibility' of the Flux of the
Renormalization Group in a 2-D Field Theory},
Pisma Zh. Eksp. Teor. Fiz. 43 (1986) 565; {\it Renormalization Group and
Perturbation Theory Near Fixed Points in Two Dimensional Field Theory,}
Yad. Fiz. 46 (1987) 1819.}

\lref\loren{N. Berkovits and O. Chand\'{\i}a, {\it
Lorentz Invariance of the Pure Spinor BRST Cohomology for the Superstring,}
Phys.Lett. B514 (2001) 394, hep-th/0105149.}

\lref\private{N. Berkovits, private communication.}

\lref\wadia{Gautam Mandal, Nemani V. Suryanarayana, Spenta R. Wadia,
{\it Aspects of Semiclassical Strings in AdS$_5$,} Phys.Lett. B543 (2002) 81;
hep-th/0206103.}



\Title{\vbox{\hbox{ IFT-P.076/2002}}}
{\vbox{\centerline{One Loop Conformal Invariance of the Superstring}
\smallskip
\centerline{in an $AdS_5 \times S^5$ Background}}}
\centerline{Brenno Carlini Vallilo\foot{E-Mail:
vallilo@ift.unesp.br}}
\bigskip
\bigskip
\centerline{\it Instituto de F\'\i sica Te\'orica, Universidade
Estadual Paulista}
\centerline{\it Rua Pamplona 145, 01405-900, S\~ao Paulo, SP, Brasil}

\vskip .3in
It is proven that the pure spinor superstring in an $AdS_5 \times S^5$
background remains conformally invariant at one loop level in the sigma model
perturbation theory.

\Date{October 2002}


\newsec{Introduction}

A consistent world sheet description of the superstring in an
$AdS_5 \times S^5$ space with Ramond-Ramond flux could be a
powerful tool to study aspects of the AdS/CFT correspondence
\adscft\adsrev. The two main formalisms for the superstring suffer
from
problems that reduce severely their utility. Berkovits has
proposed a new formalism where quantization can be performed
preserving full $D=10$ supersymmetry \berk. This formalism can
also be generalized to curved backgrounds, including RR
fluxes. In Berkovits' description, kappa symmetry of the
standard GS superstring is replaced by a BRST like symmetry,
constructed in terms of the fermionic constraints $d_\a$ and an
appropriate set of bosonic pure spinors ghosts.

The GS action in an $AdS_5 \times S^5$ space was constructed in
\metsaev. Their construction was based on the coset supergroup
$PSU(2,2|4)/SO(4,1)\times SO(5)$. Sigma models in supergroups were
considered in \bersha\ and \adswitten. In \bersha\ general aspects
of the theory based on supergroups $PSL(N|N)$ were studied. In
particular, it was shown that these models are exactly conformal.
In the work \adswitten, a full superstring theory on the
supergroup manifold $PSU(2|2)$ is defined, based on the hybrid
formalism \calabi\nv. The world sheet theory is also exactly
conformal and has a non linear $N=2$ superconformal symmetry. The
superconformal symmetry is reminiscent of the hybrid string and is
used to define physical states and correlation functions.
Quantization of sigma models in coset supergroups was done in \bz.
One loop conformal invariance of this theories could be proved
including a WZ term with a specific coefficient. Little attention
was paid to the ghosts in \bz\ because in the $AdS_2 \times S^2$
they are spacetime scalars, and their action is still free in curved
spaces.

In the present work, one loop conformal invariance of the pure
spinor superstring in an $AdS_5 \times S^5$ will be proved. The
matter contribution to the sigma model was proved in \bz\ to have
vanishing beta function. But to define a string theory, ghosts
are essential. In the case of the ten dimensional superstring, the
ghosts are not scalars and they couple to the background. It is
these couplings that are analyzed in this work.

This paper is organized as follows. In section 2 a short
introduction to Berkovits' pure spinor formalism for the
superstring is given. After that, relevant facts about the
$PSU(2,2|4)$ algebra and the $AdS_5 \times S^5$ space are
introduced. In section 3 the sigma model action obtained in
section 2 is quantized and the ghost contribution to the effective
action is calculated and is shown to have no UV divergencies. The
last section has the conclusion and discusses perspectives related to
superstrings in $AdS_5 \times S^5$.

\newsec{Pure Spinor Superstring}

In this section we give a brief review of the pure spinor formalism for the
superstring. At first we discuss the flat background, and then, the
$AdS_5 \times S^5$ space as a coset supermanifold.

In flat space, the pure spinor superstring action is given by

\eqn\flat{S=\int d^2 z [\half \p x^m \overline\p x_m +
p_\a\overline\p\t^\a + \widehat p_\ah \p\th^\ah ] + S_{\l} +
S_{\lh},} where $(x^m,\t^\a,\th^\ah)$ parameterize the
$D=10$, $N=2$ superspace and $(p_\a,p_\ah)$ are the fermionic
conjugate momenta. $S_\l$ and $S_\lh$ are the free field actions
for the bosonic left and right-moving ghosts $\l^\a$ and $\lh^\ah$
satisfying the pure spinor conditions

\eqn\pure{ \l\g^m\l=0 {\rm ~~and ~~} \lh\g^m\lh=0
{\rm ~~for~~} m=0 {\rm ~to~}  9.}

Although an explicit form  of $S_\l$ and $S_\lh$ in terms of
$(\l,\lh)$ and their conjugate momenta $(\o,\oh)$ requires
breaking $SO(9,1)$ (or its euclidean version $SO(10)$) to a
subgroup, the OPE's of $\l^\a$ and $\lh^\ah$ with their Lorentz
currents $N^{mn}=\half\o\g^{mn}\l$ and $\widehat
N^{mn}=\half\oh\g^{mn}\lh$ are manifestly SO(9,1) covariant. The
condition \pure\ implies that $\o$ and $\oh$ are defined only up
the gauge invariances

\eqn\gauge{ \d\o_\a=\L^m (\g_m \l)_\a, \quad
\d\oh_\ah=\hat \L^m (\g_m\lh)_\ah,}
for any $(\L^m, \hat \L^m)$.

The action \flat\ is invariant under the supersymmetry generated by
\eqn\susy{ q_\a = \oint [ p_\a +  (\t \gamma^m)_\a \p x_m +
{1\over{12}}(\t\gamma^m)_\a (\t\gamma_m \p \t) ] }
$$ \hat q_\ah = \oint [ \hat p_\ah + (\th\gamma^m)_\ah \bar\p x_m +
{1\over{12}}(\th\gamma^m)_\ah (\th\gamma_m \bar\p \th)].$$

It is useful to define supersymmetric operators in terms of the
free world sheet fields

\eqn\defdtwo{d_\a=p_\a - (\Pi^m -{1\over 2}\t\g^m\p\t) (\g_m\t)_\a,
\quad \Pi^m = \p x^m + \t\g^m\p\t,}
$$ \widehat d_\ah=\widehat p_\ah -(\widehat\Pi^m -
{1\over 2}\th\g^m\overline\p\th) (\g_m\th)_\ah,
\quad \widehat\Pi^m = \overline\p x^m + \th\g^m\overline\p\th,$$
which satisfy the OPE's
\eqn\ope{d_\a(y) d_\b(z)\to -2 \g^m_{\a\b} (y-z)^{-1} \Pi_m,\quad
\widehat d_\ah(\overline y) \widehat d_\bh(\overline z)\to -2
\g^m_{\ah\bh}
(\overline y-\overline z)^{-1} \widehat\Pi_m.}

\subsec{Physical Vertex Operators}

There are two types of vertex operators. The unintegrated ones are
needed to compute tree level scattering amplitudes and are
appropriate in the computation of the cohomology \cohom\massive.
Integrated vertex operators are also needed to compute amplitudes
and to know the form of the action in general backgrounds. The
BRST operators are given by \eqn\brst{Q = \oint  \l^\a d_\a {\rm
~~~and~~~} \overline Q = \oint \lh^\ah \widehat d_\ah,} where
$\l^\a$ and $\lh^\ah$ carry ghost-number $(1,0)$ and $(0,1)$
respectively. Nilpotency and anti-commutation are easily checked
using the OPE's \ope. Integrated $V$ and unintegrated $U$ vertex
operators will be physical if they are in the BRST cohomology and
have ghost-number $(0,0)$ and $(1,1)$ respectively. The two types
are related by

$$ \{ Q,[\bar Q,V]\}=\p \bar \p U.$$

The integrated massless vertex operator has the form \berk
\eqn\intver{\int d^2 z V(z,\bar z)=\int d^2 z [
h_{mn}\Pi^m\hat \Pi^n + g_{\a\bh}\p \t^\a \bar \p \th^\bh +
 \hat g_{m\ah}\Pi^m\bar\p\th^\ah + g_{m\a}\hat \Pi^m\p\t^\a +}
$$ + (d_\a + N_\a^\b D_\b)\hat\Pi^m E^\a_m +
(\hat d_\ah + \hat N_\ah^\bh\hat D_\bh)\Pi^m \hat E^\ah_m +
(d_\a + N_\a^\b D_\b)\bar \p\th^\bh E^\a_\bh +
(\hat d_\ah + \hat N_\ah^\bh\hat D_\bh)\p\t^\b \hat E^\ah_\b + $$
$$(d_\a + N_\a^\b D_\b)(\hat d_\dh + \hat N_\dh^\gh\hat D_\gh)P^{\a\dh}], $$
where the space time superfields $(h_{mn},g_{\a\bh},\hat
g_{m\ah},g_{m\a}, E^\a_m,\hat E^\ah_m,E^\a_\bh,\hat
E^\ah_\b,P^{\a\dh})$ depend on the zero modes (not derivatives) of
$(x,\t,\th)$ only, $N_\a^\b={1\over{8}}(\g_{mn})_\a^\b N^{mn}$ and
$\hat N_\ah^\bh ={1\over{8}}(\g_{mn})_\ah^\bh \hat N^{mn} $ are
the Lorentz currents of the bosonic ghosts. This vertex operator
was first proposed by Siegel \csm\ up to the ghost terms. It can
be shown that the cohomological conditions give the equations of
motion and gauge invariances of linearized supergravity.

\subsec{Action in Curved Backgrounds}

The form of the vertex operator \intver\ helps us to write the
superstring action in a curved spacetime. It is only necessary to
covariantize the spacetime indexes with respect to diffeomorphism
invariance. The result is

\eqn\curvact{S_{curv}= \int d^2 z [ \half(G_{MN}+B_{MN})\p y^M \bar\p y^N + }
$$ + d_\a E^\a_M\bar \p y^M + \half N_{mn}\pb y^M \Omega_M^{mn} +
\hat d_\ah \hat E^\ah_M \p y^M +\half \hat N_{mn} \p y^M \hat \Omega_M^{mn} +
d_\a \hat d_\gh P^{\a\gh} + $$
$$+ \hat N_{mn}d_\a C^{\a mn} +
 N_{mn}\hat d_\ah \hat C^{\ah mn}+ {1\over 4}N_{mn}\hat
N_{op}R^{mnop} + \Phi(x,\t,\th)R \quad ] + S_{\l} + S_{\lh},$$
where the superfields $(G_{MN},B_{MN},E^\a_M,\hat
E^\ah_M,\Omega_M^{mn},\hat\Omega_M^{mn}, P^{\a\gh},C^{\a mn},\hat
C^{\ah mn},R^{mnop},\Phi)$ are related to the supergravity
multiplet \sugra. The first line in \curvact\ is just the GS
action in a curved background, the second and third lines are
necessary to covariantly quantize the superstring, {\it i.e.}, 
they provide invertible propagators (containing no operators that
might have zero modes) for the fermions. $\Phi(x,\t,\th)R$ is the
Fradkin-Tseytlin term, where $\Phi$ is the compensator scalar
superfield whose lowest component is the dilaton and $R$ is the
worldsheet curvature. In curved space, $p_\a$ and $\hat p_\ah$
lost their meanings, so we treat $d_\a$ and $\hat d_\ah$ as
fundamental fields.

\subsec{Action in an $AdS_5 \times S^5$ Background}

As was shown in \metsaev\ , the  $AdS_5 \times S^5$ background can be
described by a coset supergroup element $g$ taking values in
$PSU(2,2|4)/SO(4,1)\times SO(5)$ where the supervierbein and spin
connections are given by

$$E^A_M dy^M = (g^{-1}dg)^A, $$
where $A=(\underline a,\a,\ah,[\underline a \underline b])$ and
$\underline{a}$ signifies either $a$ or $a'$ and
$[\underline{cd}]$ signifies either $[ab]$ or $[a'b']$, $a=0$ to $4$
and $a'=5$ to $9$. The non-vanishing structure constants $f_{AB}^C$
of the $PSU(2,2|4)$ algebra are\foot{In \chan\ there is a typo in the
$f^{[ef]}_{cd}$, $f^{[e'f']}_{c'd'}$ and
$f_{[\underline{cd}][\underline{ef}]}^{[\underline{gh}]}$
structure constants.}
\eqn\structure{
f_{\a\b}^{\underline{c}} =2 \g^{\underline{c}}_{\a\b},\quad
f_{\ah\bh}^{\underline{c}} =2 \g^{\underline{c}}_{\a\b},}
$$
f_{\a \bh}^{[ef]}=
f_{\bh \a}^{[ef]}=
(\g^{ef})_\a{}^\g \d_{\g\bh},\quad
f_{\a \bh}^{[e'f']}=
f_{\bh \a}^{[e'f']}= -
(\g^{e'f'})_\a{}^\g \d_{\g\bh},$$
$$f_{\a \underline{c}}^\bh
=-f_{\underline{c}\a}^\bh
=\half (\g_{\underline c})_{\a\b}
\d^{\b\bh},\quad
f_{\ah \underline{c}}^\b =
-f_{\underline{c}\ah}^\b =
-\half
(\g_{\underline c})_{\ah\bh} \d^{\b\bh},$$
$$f_{c d}^{[ef]}= \half \d_c^{[e} \d_d^{f]},
\quad f_{c' d'}^{[e'f']}= -\half \d_{c'}^{[e'} \d_{d'}^{f']},$$
$$f_{[\underline{cd}][\underline{ef}]}^{[\underline{gh}]}=\half (
\eta_{\underline{ce}}\d_{\underline{d}}^{[\underline{g}}
\d_{\underline{f}}^{\underline{h}]}
-\eta_{\underline{cf}}\d_{\underline{d}}^{[\underline{g}}
\d_{\underline{e}}^{\underline{h}]}
+\eta_{\underline{df}}\d_{\underline{c}}^{[\underline{g}}
\d_{\underline{e}}^{\underline{h}]}
-\eta_{\underline{de}}\d_{\underline{c}}^{[\underline{g}}
\d_{\underline{f}}^{\underline{h}]})$$
$$f_{[\underline{cd}] \underline{e}}^{\underline{f}} =
-f_{\underline{e} [\underline{cd}]}^{\underline{f}}= \eta_{\underline{e}
\underline{[c}} \d_{\underline d]}^{\underline{f}},\quad
f_{[\underline{cd}] \a}^{\b} =
-f_{\a [\underline{cd}]}^{\b}= \half(\g_{\underline{cd}})_\a{}^\b,\quad
f_{[\underline{cd}] \ah}^{\bh} =
-f_{\ah [\underline{cd}]}^{\bh}= \half(\g_{\underline{cd}})_\ah{}^\bh.$$

The $PSU(2,2|4)$ algebra $\CH$ has a natural decomposition \bz\
$\CH=\sum \CH_i$, $i=0$ to $3$ \eqn\dec{J_{\underline a} \in
\CH_2, \quad J_{[\underline{ab}]} \in \CH_0, \quad J_\a \in \CH_1,
\quad J_\ah \in \CH_3.}

As can be seen from the structure constants \structure\ \eqn\alge{
[\CH_i ,\CH_j ] \subset \CH_{i+j} \quad {\rm mod }~4.} The
bilinear form also respects the decomposition \eqn\bili{ \ll \CH_i
, \CH_j \rr =0\quad {\rm unless }\quad i+j=0 \quad ({\rm mod
}~4).}

Besides the superspace geometry, the background superfields
$B_{AB}$ and $P^{\a\bh}$ also have expectation values\foot{ In
\berk\ there is a mistake in the value of these fields. The origin
of this mistake is that $Ng_s$ is the flux $\int_{S^5}F_5
e^{\phi}$ and not the value of the field \private.}
\eqn\back{B_{\a\bh}=B_{\bh\a}=-\half (Ng_s)^{1\over 4}\d_{\a\bh},
\quad P^{\a\bh}={1\over (Ng_s)^{1\over 4}} \d^{\a\bh},} where $N$
is the value of the Ramond-Ramond flux, $g_s$ is the string
coupling constant and $\d_{\a\bh}=(\g^{01234})_{\a\bh}$ with 01234
being the directions of $AdS_5$.

Since we have the value of all background fields, we can plug then
to the action \curvact . Because of the term $Ng_s\d^{\a\bh}d_\a
d_\bh$, $d_\a$ and $d_\bh$ are auxiliary fields, and may be
eliminated by their equations of motion. The final result
is \chan

\eqn\adsaction{S= \int d^2 z [
\half(\eta_{cd} J^c \overline J^d +\eta_{c'd'} J^{c'} \overline
J^{d'}) + (Ng_s)^{1\over 4}\d_{\a\bh}(3 J^\bh \overline J^\a -
J^\a \overline J^\bh) }
$$+ \half(N_{\underline{cd}} \bar J^{[\underline{cd}]} +
\hat N_{\underline{cd}}J^{[\underline{cd}]})+\half( N_{cd}\widehat N^{cd}
-N_{c'd'}\widehat N^{c'd'})]  + S_\l + S_{\widehat\l}$$
where $J^A = (g^{-1} \p g)^A$ and $\overline J^A = (g^{-1} \overline\p
g)^A$ are left-invariant currents constructed from the supergroup element
$g \in PSU(2,2|4)$, $[N^{cd},N^{c'd'}]$ and
$[\widehat N^{cd},\widehat N^{c'd'}]$
are the $SO(4,1)\times SO(5)$ components of the Lorentz current
for $\l^\a$ and $\lh^\ah$, and $S_\l$ and $\widehat S_\lh$ are the
same as in \flat. Performing the rescaling\foot{The ghosts also need
to be rescaled, in order to have same weight with respect to
lorentz transformations. This implies
$(\l,\omega,\hat \l,\hat\omega)\to(Ng_s)^{1\over 8}$ and
$(N,\hat N)\to(Ng_s)^{1\over 4}$.}

\eqn\resc{J^{\underline c}\to (Ng_s)^{1\over 4},
\quad J^{\a}\to (Ng_s)^{1\over 8},
\quad J^{\ah}\to (Ng_s)^{1\over 8},
\quad J^{[\underline{cd}]}\to (Ng_s)^{1\over 4}}
and calling $\a=(\l)^{-{1\over 4}}$, $\l=Ng_s$, the action
\adsaction\ is just a sigma model action based on the coset
supergroup $PSU(2,2|4)/SO(4,1)\times SO(5)$ with coupling constant
$\a$ coupled to the bosonic ghosts plus a WZ term

\eqn\sigmamodel{
S_{AdS}={1 \over{\a^2}}\int d^2 z\half\eta_{AB}J^A \bar J^B\big|_{\CH
\setminus\CH_0}+ k S_{WZ} + }
$$ + {1\over \a^2} \int d^2 z [\half N_{\underline c \underline d}
\overline J^{[\underline c \underline d]} 
+\half\widehat N_{\underline c \underline d} J^{[\underline c \underline d]}
+\half( N_{cd}\widehat N^{cd} -N_{c'd'}\widehat N^{c'd'})]
+ {1\over \a^2}(S_\l + S_{\widehat\l}), $$
where
\eqn\wz{S_{WZ}= {1 \over{\a^2}}\int d^2 z
{1\over{2}}\eta_{\a\bh}[ J^\bh \bar J^\a - \bar J^\bh J^\a ],}
$k=\half$ and $\eta_{AB}=(\eta_{\underline{ab}},-4\d_{\a\bh},4\d_{\a\bh},
\eta_{a [ b}\eta_{c] d}, -\eta_{a' [ b'}\eta_{c'] d'} )$.
Since the dilaton is constant in this background, the Fradkin-Tseytlin
term is integrated to give the usual genus counting coupling constant.

In \chan\ it was shown that $\d_{\a\ah}\l^\a J^\ah$ and
$\d_{\a\ah}\lh^\a J^\a$ are holomorphic and anti-holomorphic
respectively. This means that $Q$ and $\bar Q$ are conserved
charges in this background. Furthermore, when acting on massless
states these charges are nilpotent and anti-commute. This proves
that $Q$ and $\bar Q$ can be used to define massless fluctuations
around the background. The problem of computing the cohomology to
arbitrary mass level is still open.

\newsec{One Loop Beta Function}

Classically, there is only one coupling constant in \sigmamodel.
If quantum effects are taken into account this picture may change.
This will be the case if the various interactions in \sigmamodel\
have different coefficients in the renormalization group flow.
Consistency of string theory requires that these coefficients are
exactly zero. The coefficients of the renormalization group,
namely, the beta functions, are calculated renormalizing UV
divergent diagrams in the effective action. If there is no
divergence at all, the beta functions are automatically zero. In
subsection 3.1, \sigmamodel\ wll be quantized and 
the results of \bz\ on the cancellation of one loop divergences 
from the
matter part of \sigmamodel\ will be summarized.
After that, it will be
shown that a new one loop divergence appears due to the interaction
\eqn\ghostint{\int d^2 z [\half N_{\underline c \underline d}
\overline J^{[\underline c \underline d]} + \half \widehat
N_{\underline c \underline d} J^{[\underline c\underline d]}].} 
Finally, in
subsection 3.3 it will be shown that this new divergence is
exactly cancelled by the interaction 
\eqn\secondghost{\int d^2 z [\half N_{cd}\widehat N^{cd}
-\half  N_{c'd'}\widehat N^{c'd'}]}
of the pure spinor ghosts.

\subsec{Sigma Model Perturbation Theory}

It is straightforward to use background field method to quantize
\sigmamodel\ as was shown in \bz. A classical background field 
$g_0$ is
chosen and the quantum fluctuations are parameterized by
$X$, $g=g_0 e^{\a X}$, where $\a$ is the sigma model coupling
constant. The quantum currents are \eqn\currents{ J = g^{-1}\p g =
e^{-\a X}J_0 e^{\a X}+e^{-\a X}\p e^{\a X},}
$$ \bar J = g^{-1}\bar \p g = e^{-\a X}\bar J_0 e^{\a X}+
e^{-\a X}\bar\p e^{\a X},$$
where $J_0=g_0^{-1}\p g_0$ and $\bar J_0=g_0^{-1}\bar\p g_0$. A similar
expansion is assumed for the ghosts
\eqn\quanghost{(\tilde\l,\tilde\omega,\tilde\lh,\tilde\oh)=
(\l_0 + \a\l,\o_0 +\a\o,\lh_0 +\a\lh,\oh_0 + \a\oh).}
I will not enter into the details of the propagator for these fields.

The action \sigmamodel\ can be written as
\eqn\currentact{S_{AdS}= {1\over \a^2}\int d^2 z [
\half \ll J_2 ,\bar J_2 \rr + {3\over 4}\ll J_3 , \bar J_1 \rr +
{1\over 4}\ll\bar J_3,J_1 \rr ] + }
$$ + {1\over \a^2}\int d^2 z [\half N_{\underline c \underline d}
\overline J^{[\underline c \underline d]} +
\half \widehat N_{\underline c \underline d} J^{[\underline c \underline d]}
+\half( N_{cd}\widehat N^{cd} -N_{c'd'}\widehat N^{c'd'})]
+ {1\over \a^2}(S_\l + S_{\widehat\l}), $$
where $J_i=J\big|_{\CH_i}$. Since the action \currentact\ has the gauge
invariance $g \to ge^h$, where $h \in \CH_0$, is possible to chose a gauge
such that $X \in \CH \setminus \CH_0$.

To compute the effective action, the currents \currents\ are
substituted in \currentact\ and then expanded in powers of $X$. The
term independent of $X$ is the classical action, and the term quadratic
in $X$ is

$$\int d^2 z \, {\rm Str}(\p X \pb X) ,$$
which defines the propagator for the quantum fields.

The effective action is the sum of all 1PI diagrams with
background currents as external lines. The ghost part is treated
in the same way. As was shown in \bz, the one loop effective
action of the matter part of the sigma model (first line of
\sigmamodel) has no divergence for $k=\pm \half$. The beta
function is calculated renormalizing the UV divergent diagrams,
{\it i.e.}, those which have two external background lines, by a
simple power counting argument. The key fact in the computation of
\bz\ is that the $PSU(2,2|4)$ algebra has vanishing dual Coxeter
number. This implies, in particular, the identity

\eqn\coxeter{f_{a\a\b}f_{\bh\ah b}\eta^{\b\bh}\eta^{\a\ah} +
f_{a\ah\bh}f_{\b\a b}\eta^{\bh\b}\eta^{\ah\a} +
2f_{a[cd]e}f_{f[gh]b}\eta^{ef}\eta^{[cd][gh]}=0.}

For example, with the help of \coxeter\ they were able to sum up
all the divergent contributions with $J^{\underline{a}}_0\bar
J^{\underline{b}}_0$ as the external background currents to get
the result

\eqn\dive{\eqalign{\eqalign{\epsfbox{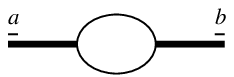}}+
\eqalign{\epsfbox{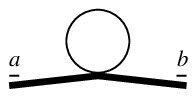}\cr \noalign{\vskip0.2in}}=
\eqalign{ 2\pi\big( \half -2 k^2 \big)
{\rm log}\big( {\L \over \mu} \big)
J^{\underline{a}}_0 \bar J^{\underline{b}}_0
\big( f_{\underline{a}[\underline{cd}]\underline{e}}
f_{\underline{f}[\underline{gh}]\underline{b}}
\eta^{\underline{ef}}\eta^{[\underline{cd}][\underline{gh}]} \big)}}.}
A similar structure occurs in the other divergent diagrams.
Therefore, for $k=\pm \half$ the order
$\a^2$ matter contribution to the
beta function of the coupling $1\over \a^2$ vanishes.

In the present work, the following identity implied by the
vanishing of the dual Coxeter number will be important:
\eqn\iden{ f_{[\underline{ab}]\a\bh}f_{\b\ah[\underline{cd}]}
\eta^{\bh\b}\eta^{\a\ah}+
f_{[\underline{ab}]\ah\b}f_{\bh\a[\underline{cd}]}\eta^{\b\bh}\eta^{\ah\a} +}
$$+f_{[\underline{ab}]\underline{ef}}f_{\underline{gh}[\underline{cd}]}
\eta^{\underline{fg}}\eta^{\underline{eh}} +
f_{[\underline{ab}][\underline{ef}][\underline{gh}]}
f_{[\underline{lm}][\underline{no}][\underline{cd}]}
\eta^{[\underline{gh}][\underline{lm}]}
\eta^{[\underline{ef}][\underline{no}]} = 0.$$

\subsec{Sigma Model Calculation}

I am interested in the ghost contribution to the effective action
that comes from the second line of \currentact. This contribution
is divided in two parts. The first is a sigma model calculation
involving the term \ghostint.
And the second part 
comes just from the ghosts in the term \secondghost.

Although $X$ is gauge fixed to be
in $\CH \setminus \CH_0$, $J\big|_{\CH_0}$ will have quantum fluctuations
given by \currents \foot{Henceforth, to avoid cumbersome notation, I will
call $J_0\big|_{\CH_0}$ just by $J_0$, since the other background currents
are not going to appear anymore.}
\eqn\quant{J\big|_{\CH_0}=J_0 + \a^2 \big( [\p X_2,X_2] + [\p X_1,X_3] +
[\p X_3,X_1]\big) + ... }
$$\bar J\big|_{\CH_0}=\bar J_0 + \a^2 \big( [\pb X_2,X_2] + [\pb X_1,X_3] +
[\pb X_3,X_1]\big) + ... $$ where $...$ means higher order terms
in $\a$ and involving other background currents, which will play
no rule here. So, the interaction has the form \eqn\inte{
\half\int d^2 z {\rm Str}\big( N_0 [\pb X_2,X_2] + N_0[\pb
X_1,X_3] + N_0 [\pb X_3,X_1] +}
$$ \hat N_0 [\p X_2,X_2] + \hat N_0 [\p X_1,X_3] +
\hat N_0 [\p X_3,X_1] \big),$$
where $N_0$ and $\hat N_0$ are the background ghost currents contracted
with the generators of $\CH_0$. At one loop these interactions will give rise
to fish type diagrams. The combined divergent contribution is
\eqn\divone{\eqalign{\eqalign{\epsfbox{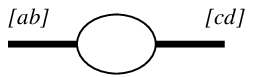} \cr
\noalign{\vskip0.5in}}\eqalign{=2\pi
{1\over 4}{\rm log}\big( {\L \over \mu} \big) N_0^{\underline{ab}}
\hat N_0^{\underline{cd}} \big( & f_{[\underline{ab}]\a\bh}
f_{\b\ah[\underline{cd}]}\eta^{\bh\b}\eta^{\a\ah}+ \cr
& f_{[\underline{ab}]\ah\b}f_{\bh\a[\underline{cd}]}\eta^{\b\bh}\eta^{\ah\a} +
\cr & f_{[\underline{ab}]\underline{ef}}f_{\underline{gh}[\underline{cd}]}
\eta^{\underline{fg}}\eta^{\underline{eh}} \big). }}}

Using the identity \iden, this expression can be rewritten as
\eqn\cox{-2\pi{1\over 4}{\rm log}\big( {\L \over \mu} \big)
N_0^{\underline{ab}} \hat N_0^{\underline{cd}}
\big(f_{[\underline{ab}][\underline{ef}][\underline{gh}]}
f_{[\underline{lm}][\underline{no}][\underline{cd}]}
\eta^{[\underline{gh}][\underline{lm}]}
\eta^{[\underline{ef}][\underline{no}]}     \big).} The group
theoretic coefficient turns out to be just the dual Coxeter number
of the group $SO(1,D-1)\times SO(D)$, $C_V=(D-2)$, with $D=5$,
times $\eta_{\underline{a}[\underline{c}}\eta_{\underline{d}]b}$.
Substituting that coefficient in \cox, one gets
\eqn\new{-2\pi{\rm
log}\big( {\L \over \mu} \big)(D-2) [\half N_0^{ab}\hat N_0^{cd}+
\half N_0^{a'b'}\hat N_0^{c'd'}] .} This is the only divergent
diagram coming from \ghostint. It can be shown that the analogous
expression of \inte\ containing ghost quantum fluctuations
\eqn\notcontribu{\half\int d^2 z [ \half
\bar J^{\underline{ab}}_0(\o\gamma_{\underline{ab}}\l) + \half 
J^{\underline{ab}}_0(\lh\gamma_{\underline{ab}}\oh)],} does not
contribute to the beta function to this order, since there is no
mixing between $(\l,\o)$ and $(\lh,\oh)$.

\subsec{Ghost Contribution}

The pure ghost contribution can be calculated by simple conformal
field theory technics\foot{The result in the previous subsection
could also be derived in the same way. The ghost part will be done
in a different way to avoid subtitles related to the propagators
of the ghosts.}. As was discussed in the section 2, it is
difficult to write a covariant free field action for the ghosts.
Nevertheless, the OPE's between the Lorentz generators and the
ghosts $(\l,\o)$ is covariant. Also, the Lorentz algebra of these
currents can be written in a covariant way \berk\loren

\eqn\cov{N^{\underline{ab}}(z)N^{\underline{cd}}(w) \to
{{\eta^{\underline{a}[\underline{c}}N^{\underline{d}]\underline{b}}(w) -
\eta^{\underline{b}[\underline{c}}N^{\underline{d}]\underline{a}}(w)}
\over(z-w)}
-3{{\eta^{\underline{a}[\underline{c}}\eta^{\underline{d}]\underline{b}}}
\over {(z-w)^2}},}
$$\hat N^{\underline{ab}}(\bar z) \hat N^{\underline{cd}}(\bar w) \to
{{\eta^{\underline{a}[\underline{c}}\hat
N^{\underline{d}]\underline{b}} (\bar w) -
\eta^{\underline{b}[\underline{c}}\hat
N^{\underline{d}]\underline{a}}(\bar w)} \over(\bar z-\bar w)}
-3{{\eta^{\underline{a}[\underline{c}}\eta^{\underline{d}]\underline{b}}}
\over {(\bar z-\bar w)^2}},$$ the double pole of this OPE's is
what characterizes the pure spinor nature of the Lorentz currents.
Note that the first terms of these OPE's can be calculated using
the naive OPE's for the ghosts \eqn\ghostprop{\o_\a(z)\l^\b(w)\to
{\d_\a^\b \over{(z-w)}},\quad \oh_\ah(\bar z)\lh^\bh(\bar w)\to
{\d_\ah^\bh \over{(\bar z- \bar w)}}.} It should be stressed that
these OPE's can be used because $(\l,\o)$ appear in the combination
$\o\gamma^{\underline{ab}}\l$ and no trace over the spinor indexes
is performed. That is why \ghostprop\ cannot be used to calculate
the last terms in \cov.

In the $AdS_5 \times S^5$ space, the free field theory of the ghosts becomes
interacting. The pure ghost contribution to the beta function comes from
the marginal operators
(marginal because they have conformal weight $(1,1)$ in the free theory)

\eqn\marg{\CO_1(z,\bar z)=\half N^{ab}\hat N_{ab}, \quad
\CO_2(z,\bar z)=\half N^{a'b'}\hat N_{a'b'}.} Using the naive
OPE's \ghostprop\ and basic commutation relation of the
gamma matrices,
\eqn\gammaalg{[\gamma^{\underline{ab}},\gamma^{\underline{cd}}]_\a^\b
[\gamma_{\underline{ab}},\gamma_{\underline{cd}}]_\ah^\bh =
16(D-2)(\gamma^{\underline{ab}})_\a^\b
(\gamma_{\underline{ab}})_\ah^\bh,}

$$[\gamma^{ab},\gamma^{a'b'}]=0,$$
where $D=5$, it can be shown that they satisfy the algebra
\eqn\algebra{\CO_1(z,\bar z)\CO_1(w,\bar w) \to
 {2(D-2)\over |z-w|^2}\CO_1(w,\bar w)+ ... , \quad
\CO_2(z,\bar z)\CO_2(w,\bar w)\to {2(D-2)\over |z-w|^2}\CO_2(w,\bar w)+... ,}
$$\CO_1(z,\bar z)\CO_2(w,\bar w)\to 0,$$
where $...$ is the nonmarginal part of the OPE's, which contains
operators that need an explicit parametrization of the pure
spinors to be computed.

At one loop\foot{The nonmarginal part of the OPE in \algebra\
contribute to higher loops, where the pure spinor character will
be important.} these interactions generate divergences in the
effective action coming from
$$ \half \int d^2z \int d^2w [ \CO_1(z,\bar z)\CO_1(w,\bar w) +
\CO_2(z,\bar z)\CO_2(w,\bar w)]$$

given by \zamo
\eqn\fin{S_{div} = 2\pi {\rm log}\big( {\L \over \mu} \big)
(D-2) \int d^2z[\CO_1 + \CO_2],}
which exactly cancels the sigma model part.

\newsec{Concluding Remarks and Perspectives}

In this paper it was shown that the sigma model action of the pure
spinor superstring is conformally invariant at one loop level in
perturbation theory. Note that the proof presented here also applies to the
hybrid superstring description in an $AdS_3 \times S^3$ background
of \six, after taking into account the corrections in the
footnotes of \berk\ and the curvature coupling to the ghosts.
The calculation is important since it shows that the action
\sigmamodel\ is a good starting point to quantize the superstring
in the $AdS_5 \times S^5$ space. I say starting point because more
work has to be done.

For example, in \chan\ it was checked classically
that the BRST currents $\d_{\a\ah}\l^\a J^\ah$ and
$\d_{\a\ah}\lh^\ah \bar J^\a$ are holomorphic and give charges
that are nilpotent and anti-commute. These facts need to be
checked quantum mechanically.

It would also be interesting to check that
the conformal algebra of the energy
momentum tensors
\eqn\stress{T= \half\eta_{\underline{ab}}J^{\underline a}J^{\underline b}
- 4\d_{\a\bh}J^\a J^\bh + N_{\underline{cd}}J^{[\underline{cd}]} + T_\l,}
$$\bar T=\half\eta_{\underline{ab}}\bar J^{\underline a}\bar J^{\underline b}
-4\d_{\a\bh}\bar J^\a \bar J^\bh + \hat N_{\underline{cd}}\bar
J^{[\underline{cd}]} + \bar T_\lh$$ is preserved when quantum
corrections are taken into account. Together with the conformal
algebra it would be interesting to calculate the algebra of the
quantum currents \currents\ and find the non-holomorphic
corrections not implied by the group structure. This result could
be useful to calculate string amplitudes in the manner of \val.
Another interesting direction is to investigate the existence of
the infinitely many conserved (non local) currents in the
semiclassical limit found in \wadia.

\vskip 15pt {\bf Acknowledgements:} I would like to thank Daniel
Nedel for useful discussions and especially Nathan Berkovits for
useful discussions and suggestions, and FAPESP grant 00/02230-3
for financial support. I also thank the California Institute of
Technology, where parts of this work have been done, for
hospitality.

\listrefs
\end